# VADER:
# a VAriable Dose-rate External $^{137}$Cs irradiatoR
# for
# internal emitter and low dose rate studies

Guy Garty[1,2,*], Yanping Xu[1,†], Gary W. Johnson[2], Lubomir B. Smilenov[2], Simon K. Joseph[3], Monica Pujol-Canadell[2], Helen C. Turner[2], Shanaz A Ghandhi[2], Qi Wang[2], Rompin Shih[4], Robert Morton[2], David Cuniberti[2], Shad R. Morton[2], Carlos Bueno-Beti[2] , Thomas L. Morgan[5], Peter F. Caracappa[5], Evangelia C.Laiakis[6,7], Albert J. Fornace Jr[6,7], Sally A. Amundson[2], David J. Brenner[2]

[1]Radiological Research Accelerator Facility, Columbia University, Irvington, NY, 10533, USA

[2]Center for Radiological Research, Columbia University, New York, NY 10032, USA

[3]David A. Gardner PET Imaging Research Center, Columbia University, New York, NY, 10032, USA

[4]Department of Radiation Oncology, Columbia University, New York, NY 10032, USA

[5]Environmental Health and Safety, Columbia University, New York, NY, 10032, USA

[6]Department of Oncology, Lombardi Comprehensive Cancer Center, Georgetown University, Washington DC, 20057, US

[7]Department of Biochemistry and Molecular & Cellular Biology, Georgetown University, Washington DC, 20057, USA

[*]corresponding author: gyg2101@cumc.columbia.edu

[†] Deceased

**ABSTRACT**

Beyond prompt irradiation, $^{137}$Cs is likely to be the most biologically important agent released in many accidental (or malicious) radiation exposure scenarios. $^{137}$Cs either can enter the food chain or be consumed or if present in the environment (e.g. fallout) can provide external irradiation. In either case, due to the high penetration of the 662 keV γ rays emitted by $^{137}$Cs, the individual will be exposed to a uniform, whole body, irradiation at low dose rates.

The VADER (VAriable Dose-rate External $^{137}$Cs irradiatoR) allows modeling these exposures, bypassing many of the problems inherent in internal emitter studies. Making use of discarded $^{137}$Cs brachytherapy seeds, the VADER can provide varying low dose rate irradiations at dose rates of 0.1 to 1.2 Gy/day. The VADER includes a mouse "hotel", designed to allow long term simultaneous residency of up to 15 mice. Two source platters containing ~250 mCi each of $^{137}$Cs brachytherapy seeds are mounted above and below the cage and can be moved under computer control to provide constant low dose rate or a varying dose rate mimicking $^{137}$Cs biokinetics in mouse or man. We present the VADER design and characterization of its performance.

# INTRODUCTION

In many large-scale radiation exposure scenarios, internal or external exposure to $^{137}$Cs is often the major source of radiation exposure [1-3]. $^{137}$Cs is generated in large quantities by fission and is one of the major sources of contamination following reactor accidents [3] and in nuclear fallout, for ground burst nuclear detonations [4] but not for air burst scenarios [5]. Due to its ubiquity in industrial irradiation systems, it presents a hazard for both malicious and accidental exposures (similar to the Goiânia accident [6]).

$^{137}$Cs is a water-soluble γ emitter with a physical half-life of 30 years. When released, it can readily be deposited on surfaces and enter the food chain [7, 8]. Because of its environmental persistence and ease of dispersal, $^{137}$Cs poses a significant risk to the general public. It is therefore important to develop biological assays for identifying individuals exposed to internal or environmental $^{137}$Cs. With external exposures, following an improvised nuclear device (IND) for example, dose rates may be initially high, depending on the level of shielding and distance from the detonation [9], and decrease rapidly with time [10], following the so called 7-10 rule [11]. Typical exposure durations are on the order of 48 h to 1 week, depending on the time required for evacuation, and doses of multiple Gy are possible [9].

In the case of internal exposures, dose rates are typically much lower [4] but the persistence may be much higher: $^{137}$Cs permeates all tissues and is cleared over a period of months [12] (although chelating agents somewhat accelerate its clearance [13]), accumulating damage to tissues and particularly to the genetic material, which may result in delayed disease. Thus, in the event of a large-scale accidental or malicious release of volatile radionuclides, resulting in internal or external exposures, there is an important need to develop radiation biodosimetry assay(s) and technologies for population-based triage and subsequent dose-dependent medical management [2].

Over the past decade, we have utilized $^{137}$CsCl$_{(aq)}$ injection in mice to study the long-term effects of ingested $^{137}$Cs on cytogenetic [14, 15], transcriptomic [16] and metabolomic [17, 18] endpoints. The difficulty of these studies lies in the fact that both the study animals and the resulting biofluids are often radioactive and require dedicated equipment to analyze, making systematic studies difficult and expensive. At this time, only one or two laboratories in the US have the appropriate facilities and highly trained personnel that are required for conducting these $^{137}$Cs injection studies.

As the majority of dose delivered from this type of exposure is due to long range γ rays [19] (Half Value Layer : 82 mm in water [20]), the physical dose distribution will be very similar for internal *vs.* external exposures. Thus, it is possible to model internal exposures (e.g. from ingested fallout) using an external isotope source, provided it is properly modulated to model biokinetics [21].

We have developed the VADER (VAriable Dose-rate External $^{137}$Cs irradiatoR) to facilitate modeling of low dose rate $^{137}$Cs exposures either from ingested or fallout exposures. The VADER irradiator is based on "repurposing" of old $^{137}$Cs brachytherapy seeds. These seeds were much used starting in the 1980s to treat cervical cancer at low dose rate [22], but are no longer in use clinically and so the seeds are kept in long-term storage, making them available for research use. At Columbia University, we had available a few dozen such sources, ranging from 3 to 35 mCi each. We describe below the construction and first tests of the VADER system, culminating in several 30-day mouse irradiations, mimicking the internal exposure experiments described in [15-18].

# VADER DESIGN

## Overview

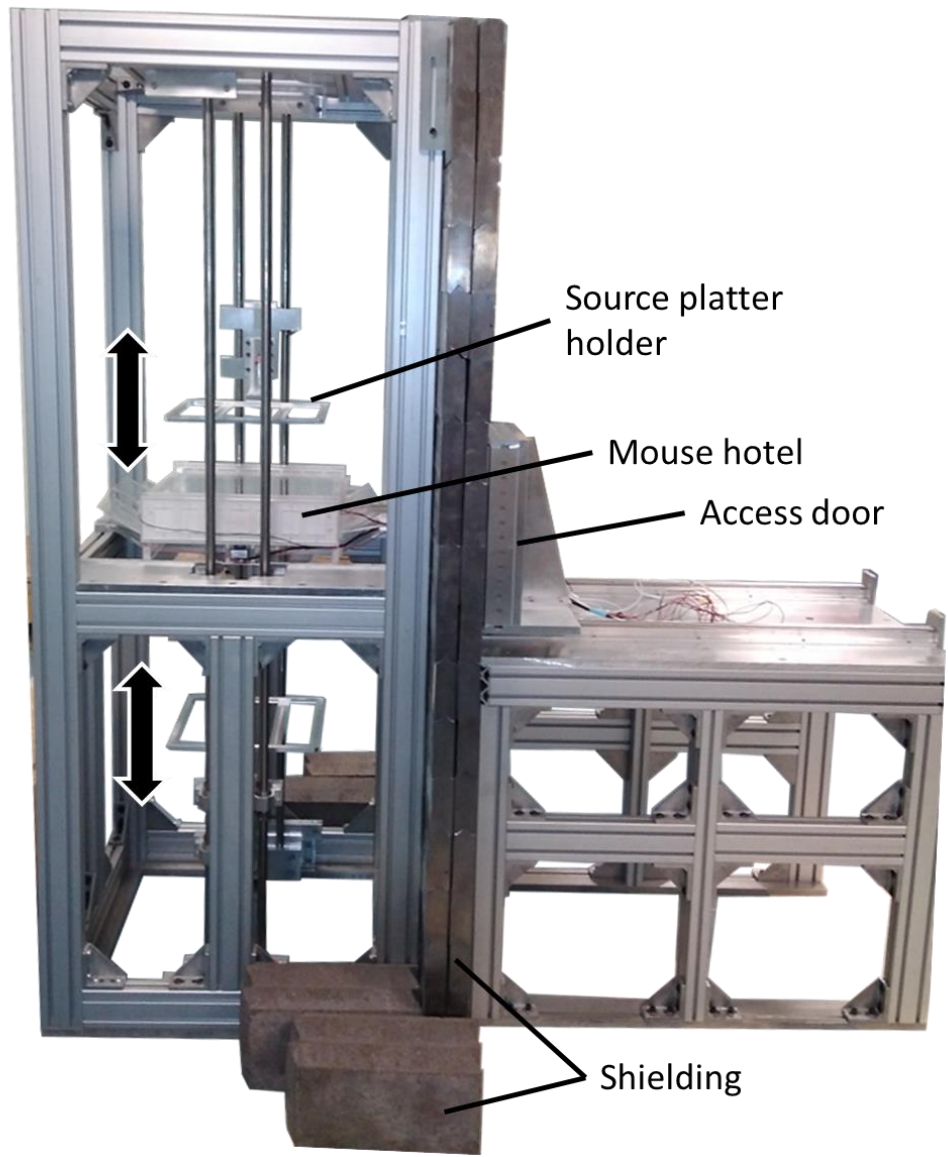

*Fig 1. The VADER structure before installation of the shielding*

The VADER system (shown in Figure 1) was designed around the requirement to house 15 mice at a time in an environment with dose rates of up to approximately 1 Gy/day over a period of several weeks. The VADER geometry was therefore dictated by the size of the mouse "hotel". Within the hotel, mice are free to move around, eat and drink *ad libitum*. Temperature, humidity, airflow and lighting are fully controlled to the required animal care standards. Shielding is provided by a mixture of lead and high-density concrete bricks as described below.

Approximately 0.5 Ci of $^{137}$Cs brachytherapy seeds, housed in two platters (above and below the mouse hotel), can be moved from almost touching the mouse hotel to about 60 cm above and below it, allowing a time-variable dose rate to be implemented. Arrangement of the sources within the platter was optimized to achieve as uniform as possible a radiation field at about 1 Gy/day.

A feature of the VADER is that there are no fixed active components within the VADER enclosure. All environmental controls and monitoring are integrated into the removable mouse hotel so that they can be easily replaced in case of radiation damage. This is significantly cheaper and easier than using radiation hardened sensors and electronics. Similarly, the motors driving the source holders are placed outside the shielding and the platters are moved using steel cables and pulleys.

**Design of the mouse hotel**

A custom mouse hotel (Fig. 2) was designed with the goal of housing up to 15 mice for 5-7 days at a time (with longer residencies requiring cage cleaning and replenishment of food and water). The hotel therefore consists of a 35 cm x 35 cm x 12 cm acrylic box in which the mice are free to move and interact with each other. Mice are able to eat and drink *ad libitum* from four all-plastic, custom-made water bottles and reservoirs holding feed pellets placed behind glass rods. Sufficient bedding material is also provided within the mouse hotel. Real time monitoring of the mice is performed using a 180° fisheye USB camera (ELP, Amazon) embedded into one of the hotel walls.

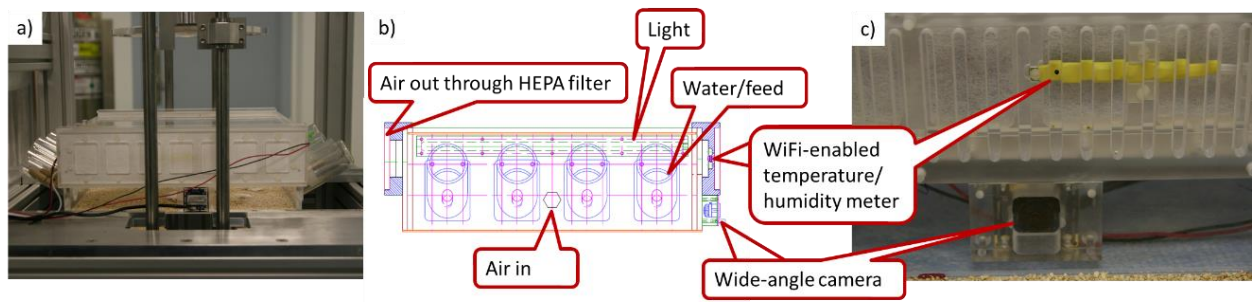

*Figure 2: a) Photo and b) design drawing of the mouse hotel. Air inlet and outlet, feeders, illumination, temperature/humidity sensor and camera are indicated. The latter two are shown in close up in panel c).*

Four mouse hotels were built: to allow two to be used simultaneously (one inside the VADER, and one for the zero dose control mice) and two spares to allow the weekly rapid transfer of the mice into a clean hotel with fresh bedding and replenished food and water.

To monitor the environment in the mouse hotel, a temperature/humidity sensor (HWg HTemp, TruePath Technologies Victor, NY) is integrated into one of the hotel walls. Medical grade air is piped into the hotel at a rate of 1.5 liter/min (10 volume changes/h). Used air is vented into the VADER enclosure through two 100 cm$^2$ HEPA filters (M-Bar Filter, Allentown Inc, Allentown, NJ). A day-night light cycle is maintained by integrating Warm White (3100K) LEDs (SMD3528; LEDWholesalers, Hayward, CA) into the hotel walls. The LEDs are powered using an adjustable benchtop power supply (Laskar Electronics, Erie, PA), plugged into a Tork timer (Grainger, Lake Forest, IL). The timer was set to "on" between 8 AM and 8 PM and "off" between 8 PM and 8 AM. Light intensity was adjusted to 60 lux.

**Design of the irradiation system**

When designing the source arrangement our goal was to achieve a maximal dose of 1 Gy/day (we can provide higher dose rates using a highly filtered X-ray machine [23]) with a uniformity of

±10% or better across the mouse hotel. As the VADER system is based on repurposing of existing $^{137}$Cs brachytherapy seeds, we were limited to use available sources, which vary in age, intensity and physical dimensions. The seeds available to us ranged between 3 and 35 mCi, as verified using a well counter (HDR 1000 plus, Standard Imaging Inc, Middleton, WI), and were either 20.0 or 21.0 mm long and between 3.1 and 3.2 mm in diameter.

To model various source geometries and select the seed intensities and positions, we calculated the expected dose across the mouse hotel using a MATLAB script. We modeled each seed as two point sources 7 mm apart. The dose rate at a distance *R* from each half seed is given by

$$Dose\ rate\ [Gy/day] = 67.92 \frac{\frac{1}{2} Seed\ activity\ [mCi]}{R[mm]^2}$$

Where 67.92 is the dose rate in Gy/day, 1 mm away from a 1 mCi point source, obtained using the RadPro Calculator (http://www.radprocalculator.com/Gamma.aspx) and a quality factor of 1.

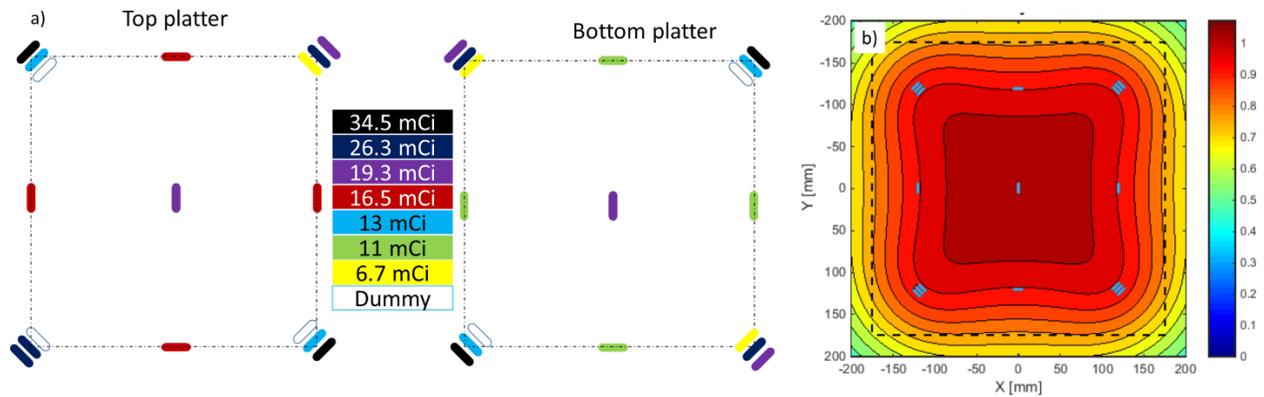

*Figure 3: a) The source arrangement used in VADER. b) MATLAB simulation of the dose rate obtained from this arrangement. The dashed line corresponds to the extents of the mouse hotel; each contour corresponds to a 5% variation in expected dose rate.*

The ideal source placement above and below the mouse hotel, within the geometric and source availability constraints is shown in Fig. 3. Eight sets of three 20 mm long seeds, totaling about 50 mCi per set were placed above/below the four corners of the mouse cage and single 21 mm long 10-17 mCi seeds at the centers of the edges and at the center of the cage. Using this geometry, the dose rate within the mouse cage was calculated to be uniform to within 20% at the highest nominal dose rate (1 Gy/day).

Custom aluminum platters were designed to hold the seeds. Each platter (Fig. 4) consists of two layers of aluminum (3/16" and 1/8" thick) with pockets hollowed out for either three 20 mm long seeds (at the corners) or one 21 mm long seed (at the edge and platter centers). Grouping seeds by size in this way simplified the platter design and allowed tighter tolerances on the pocket lengths, preventing the seeds from shifting during use. Openings were made on both pieces so that while the seeds are captive, there is no material between the active portion of the seed and the mouse cage.

The seeds were loaded into the platters using remote manipulators at Columbia University's David A. Gardner PET Imaging Research Center. Once the seeds were loaded into the platter and the platter lid connected, the platters were loaded into a lead safe (Marshield, Burlington, ON) and transported to the VADER location. Both loading the platters into the safe and transferring the platters into the VADER were achieved using a 5' long rod with a locking gripper, designed for this purpose, and supervised by Columbia University's Radiation Safety Officer.

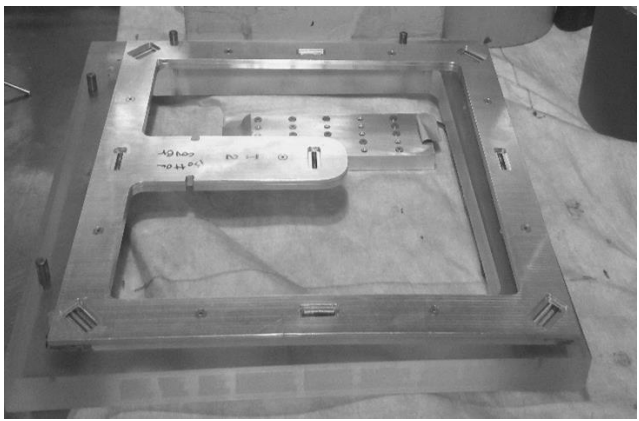

*Figure 4: The seed holder plate with seeds loaded at the hot cell.*

Within the VADER, the seed platters sit on a frame that is free to move along a vertical Steel rod, using linear ball bearings. Two Motors (AZM46MC-TS10, Oriental Motors USA Corp., Torrance, CA) are mounted outside the shielding, above the VADER. Each motor controls a spool of 3/32" ultra-flexible stainless steel wire rope (150 lb capacity, McMaster-Carr, Princeton, NJ) used to raise and lower the source platters.

The stepper motors include an absolute mechanical encoder, allowing repeatable absolute positioning, without battery back-up or external sensors. The motors also include an electromagnetic brake, which activates when power is removed. This allows the source platter to stay in place (rather than falling) in the case of a power failure. A built in 10:1 Taper hobbed gearbox provides a torque of up to 2 Nm and a motion of 0.036 °/Pulse (17 cm/turn). Each motor is independently controlled by an AZD-AD Motor driver (Oriental Motors). The two drivers are daisy-chained and controlled by RS485 Modbus.

A second steel cable connects each platter to an external weight and allows external verification of the source position, independent of the motor encoder. The motor encoders were configured such that they read zero (home) at the fully retracted position. This allows for a rapid retraction of the sources using the controller's ZHOME (fast return to home) command.

**Shielding design**

Minimizing radiation exposure to the operators and other uninvolved personnel is a prime concern in the design of $^{137}$Cs-based irradiators. Shielding for the VADER was designed to maintain radiation doses to occupationally exposed personnel (operators) in the room below 0.1 mGy/wk and 0.02 mGy/wk to anyone outside the room, in accordance with the guidance of NCRP Report 151 [24]. This also allows non-irradiated, control, mice to be housed in the same room.

The room where VADER is installed is below grade, such that the rear wall is the building's perimeter foundation wall. Adjacent rooms to the east and west and above are uncontrolled areas and considered occupied by uninvolved personnel (i.e., general public). An uncontrolled corridor borders the south side of the room. The walls are constructed of normal density concrete, 6" thick; the floor above is 4" of normal concrete.

In consultation with Columbia University's Radiation Safety Office, we have determined the required shielding to be either 3" of lead or 12" of high density concrete (4.7 g/cm$^3$; HDC). The side walls of VADER were built of two layers of 6"x6"x12" HDC interlocking bricks (Ultraray Inc, Oakville, ON, seen at the bottom of Fig. 1). Half bricks were used to offset the vertical seam every other row, to increase structural stability. An external aluminum frame was provided to increase lateral stability of the walls. The front and top faces of the VADER, however, required penetrations for inserting/removing the mouse hotel as well as for control cables and were therefore made of lead, which is much more expensive but can be machined. We used 1.5"x4"x12" interlocking lead bricks (Radiation Protection Products Inc. Wayzata, MN), similarly arranged in two layers with half bricks used to offset the seams. A custom sliding door made of a 3" thick lead slab, held in an aluminum frame was designed and built, to allow access to the mouse hotel. Although heavy, the door slides on rails and can be easily opened and closed with one hand. The rear face of the irradiator was not shielded as it abuts the north wall, is below grade and hence, there are no occupied spaces beyond.

**VADER control and monitoring**

Control software for the VADER was written in Visual C$^{++}$ (Microsoft, Redmond, WA). The software consists of a single windows "form" that controls the source movement and camera as well as monitoring the temperature and humidity in the mouse hotel. The PC running the VADER control software was configured to allow for secure remote access to allow monitoring of VADER operations.

As an experiment could last 30 days or more, it is crucial to allow recovery from unforeseen computer reboots (e.g. system updates or power failures). The VADER control software was therefore configured to automatically start at computer start up (prior to login) and maintain files containing the start time of irradiation and the required irradiation profile: a list of required source positions as a function of time. During operation, the control software periodically compares the elapsed time since the beginning of the experiment to this list, and verifies that the positions reported by the motors correspond to the required ones, moving the sources if necessary. The motor position and status as well as the temperature and humidity probe values, are displayed on screen and saved to a comma separated variable (csv) log file, which can be imported into excel.

## DOSIMETRY

Offline dosimetry of the VADER was performed using an ionization chamber (10x6-6, Radcal Corp., Monrovia, CA). Following a background measurement outside the VADER the ionization chamber and electrometer were placed at the center of the mouse hotel and dose integrated for 5 minutes at each retraction position. The required retraction as a function of dose rate was interpolated by fitting to a third order polynomial (Fig. 5). When monitoring VADER operations, the nominal dose rate was calculated based on the source position and ionization chamber results.

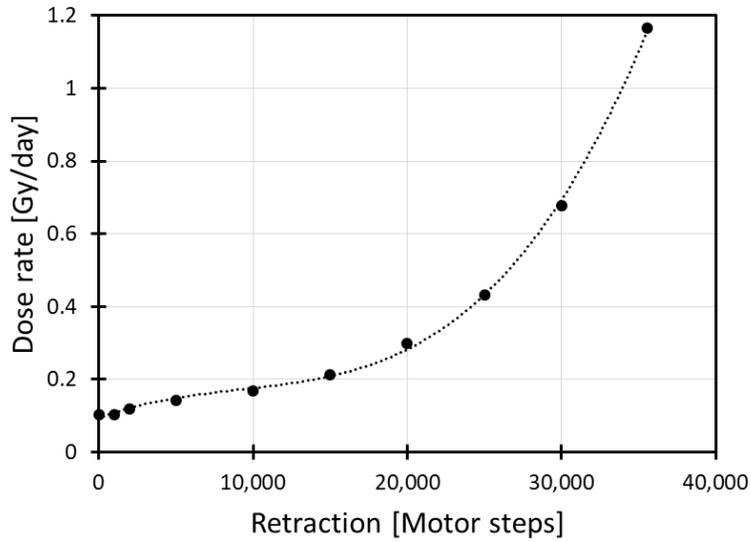

*Figure 5: Dose rate as a function of reported motor position, measured using an ionization chamber. The dotted line is the fitted curve.*

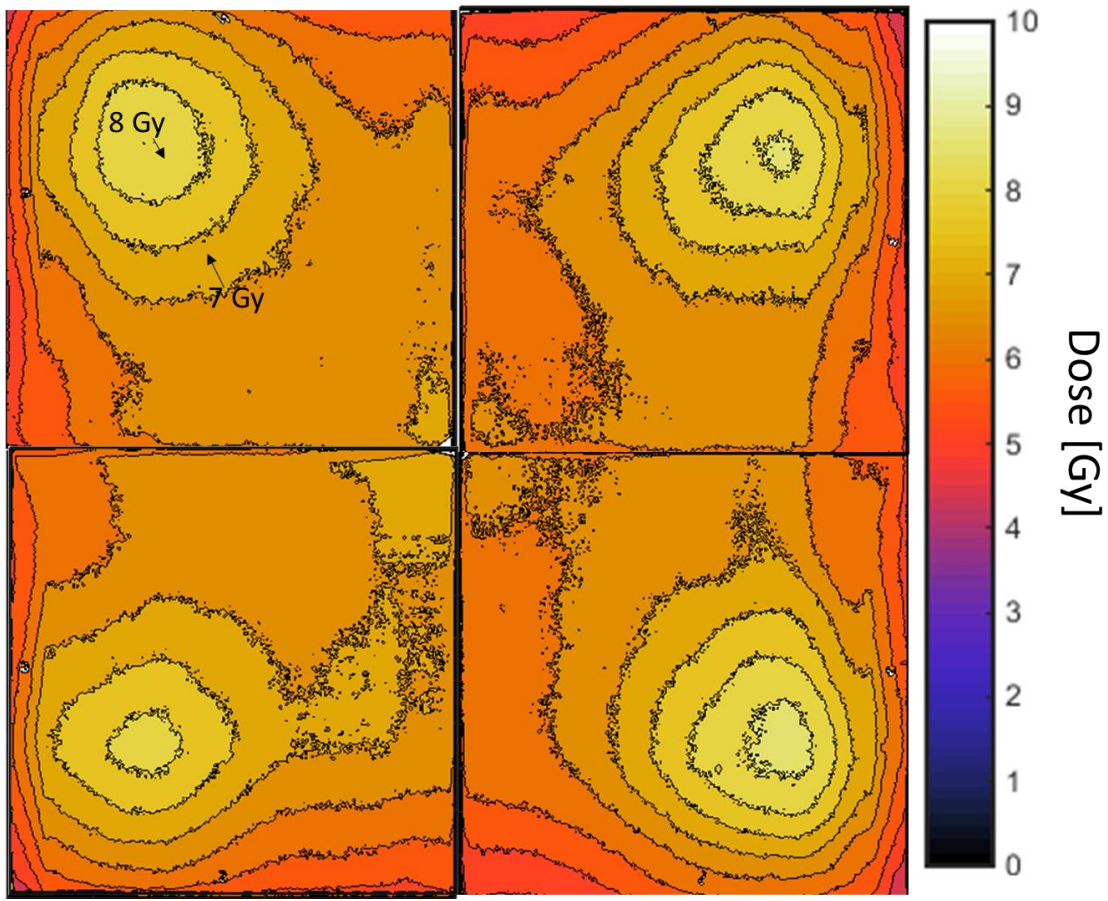

*Figure 6: Field uniformity, as measured using Gafchromic film at the highest dose rate (1.2 Gy/day). The four panels correspond to four 17.5cm square films exposed simultaneously, covering the entire mouse hotel area. Contours correspond to 0.5 Gy (roughly 5%) steps.*

Dose uniformity was evaluated using EBT3 radiochromic film (Ashland Advanced Materials, Bridgewater, USA). Four 17.5cm square pieces of film were taped down at the bottom of the mouse hotel and exposed for several days. Films were scanned on a 48 Bit RGB Epson Perfection V700 Photo flatbed scanner (Epson, Japan), with a resolution of 300 dpi. We used the web application radiochromic.com [25] to analyze the films. The program used a set of calibration films, obtained by exposing film to known doses of 250 kVp x-rays, to define a relationship between the optical density of the three colors and a given dose. The two-dimensional dose distribution for a given film could then be determined using this relationship with high accuracy.

Figure 6 shows the dose uniformity across the surface are of the mouse hotel at the highest dose rate. Variation of measured dose across the film is 10% (standard deviation). Hot spots are evident near the 50 mCi source clusters at the highest dose rates. These average out as the sources retract from the cage at 1 Gy/day the standard deviation across the cage was reduced to 3.5%.

*In vivo* dosimetry was performed on a mouse-by-mouse basis, by injecting a glass encapsulated TLD chip into each mouse [26]. TLD rods were then read using a Harshaw 2500 TLD reader (Thermo Fisher Scientific), most experiments used a heating profile consisting of a 5 °C/sec ramp up to 300 °C followed by a short hold at 300 °C and cool down to 50 °C. Dose was reconstructed based on the integrated light yield at a temperature higher than 180 °C to eliminate the low temperature, time dependent, glow peak.

## MOUSE STUDIES

All animal experiments were approved by the Columbia University Institutional Animal Care and Use Committee (IACUC; approved protocol AAQ2410) and were conducted under all relevant federal and state guidelines.

Before starting any biological study for biodosimetry endpoints, we placed 15 mice into the mouse hotel and monitored them over 2 weeks. No weight loss and normal food and water consumption was found. Initially, the mice were exploratory in the new environment and after about 30 minutes settled down. Similar behavior was observed in the VADER experiments.

### Mouse handling

For a typical experiment, 30 (15 irradiated and 15 control, randomly assigned) 7 week old C57BL/6J "cage mate" mice were purchased from Charles River Laboratories (Frederick, MD) and kept at the Columbia University Irving Medical Center animal facility for one week of adaptation. Glass-encapsulated TLD rods [26] were injected at 8 weeks of age as follows:

> Anesthesia was induced with 2% isoflurane delivered in 100% oxygen for <3 min before the implantation procedure. The encapsulated TLD rods (one per mouse) were placed in a 12-gauge needle coupled with a needle injector (Allflex, Irving, TX) and administered by subcutaneous injection in the dorsal neck. Following implantation mice were monitored up to 48 hours for complications.

Animals were then placed in the VADER, provided with food and water *ad libitum* and kept on a 12:12 hour light-dark schedule. For exposures longer than 1 week, the mouse hotel was removed from the VADER and mice transferred to a clean hotel with fresh bedding, food and water every 5-7 days. During this transfer, mice were outside the irradiator for a total of 5 min. Control mice were placed in an identical cage in the same room and were transferred to a clean hotel roughly an hour before the irradiated mice. Temperature and humidity were monitored in both hotels.

**Cell counts**

Peripheral whole blood samples were collected at euthanasia from each mouse by cardiac puncture using a heparin-coated syringe. The spleen was excised and homogenized and the isolated splenocytes were suspended in PBS 2% FBS. Cell counts were determined by flow cytometry (CytoFLEX, Beckman Coulter, Pasedena, CA) using 20 µl of heparinized blood or 20 µl of spleen suspension stained with mouse CD45 antibody (WBC marker, clone 30-F11, eBioscience, San Diego, CA).

**Apoptosis**

Peripheral blood cells and splenocytes were fixed using the FIX & PERM™ Cell Permeabilization Kit (Thermo Fisher Scientific™, Waltham, MA) and stained with nuclear dye DRAQ5 (Thermo Fisher Scientific™, Waltham, MA). The cells were measured using the ImageStream®X Mark II Imaging Flow Cytometer (LUMINEX Corporation, Austin, Texas). Images of 5000 cells per sample were acquired at 40X magnification using the 488 nm excitation laser. Captured images were analyzed using IDEAS® (LUMINEX Corporation, Austin, Texas) software for measuring proportion of apoptotic cells, as identified by automated image analysis based on nuclear imagery features in combination with bright field morphology [27]. Data was expressed as mean and standard deviation and were analyzed using GraphPad Prism 6.0 (GraphPad Software, San Diego, CA, USA). The Kruskal-Wallis test was performed to compare data among all study groups. The Mann-Whitney *U* test was used to compare between two groups. P < 0.05 was considered statistically significant.

**One Week Study**

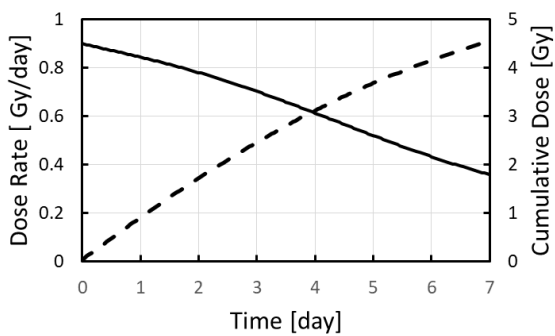

*Figure 7: VADER programming. The solid curve (left axis) displays the target dose rate as a function of time. The dashed curve (right axis) displays the target cumulative dose as a function of time.*

*Table 1: Dose rates and doses for the one week study.*

| Day | Dose rate [Gy/day] | Nominal Dose [Gy] | Average TLD dose [Gy] |
|---|---|---|---|
| 1 | 0.84 | 0.87 | 0.66±0.03 |
| 4 | 0.6 | 3.1 | 2.5±0.3 |
| 7 | 0.36 | 4.5 | 4.33±0.14 |

The first VADER experiment focused on system validation. Twelve mice were housed in the VADER mouse hotel (with three controls in a second hotel, in the same room). The VADER was programmed to produce the dose profile show in Figure 7 and Table 1, mimicking the dose and dose rate kinetics after injection of 6.66MBq $^{137}$CsCl$_{(aq)}$ [15]). Four irradiated and one control mouse were sacrificed on days 1, 4 and 7. Blood leukocyte and splenocyte cell counts (CD45+ cells) are shown in Figure 8. As expected, the cell counts decreased with increasing absorbed dose over the irradiation period. Due to the large variation in the control mice, only the day 7 leukocyte and splenocyte counts were significantly lower than the control (p<0.05).

The apoptosis frequency in the mouse peripheral blood leukocytes and splenocytes, respectively is shown in Figure 9. Although the overall rate of formation of apoptotic cells was relatively low (<3%), there was a significant (p<0.05) increase in apoptotic blood leukocytes by day 7 (dose = 4.33 Gy) and in splenocytes by day 4 (dose = 2.5 Gy) and a further increase by day 7.

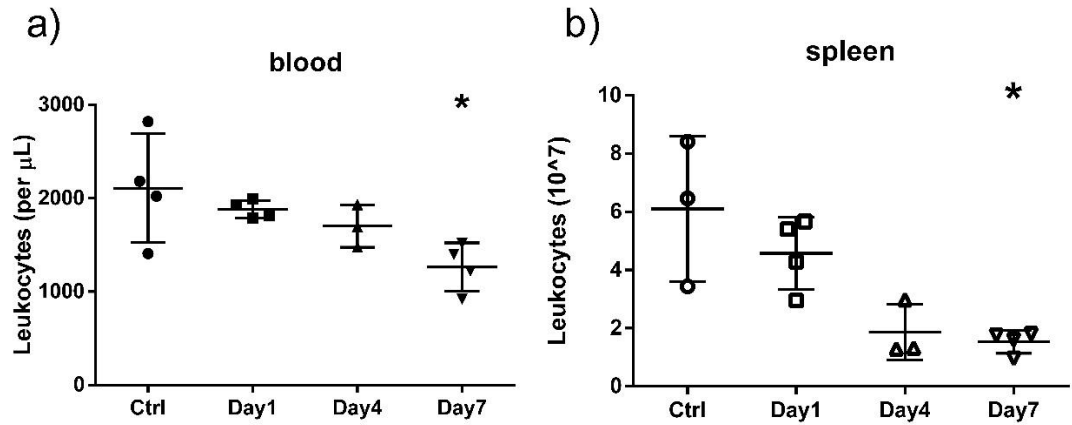

*Figure 8: Leukocyte (CD45+) counts in a) peripheral blood and b) spleen, following irradiation using the protocol shown in Figure 7. Symbols correspond to individual mice, the center line and error bars are average and Standard deviation. Asterisk denotes p<0.05 compared to control.*

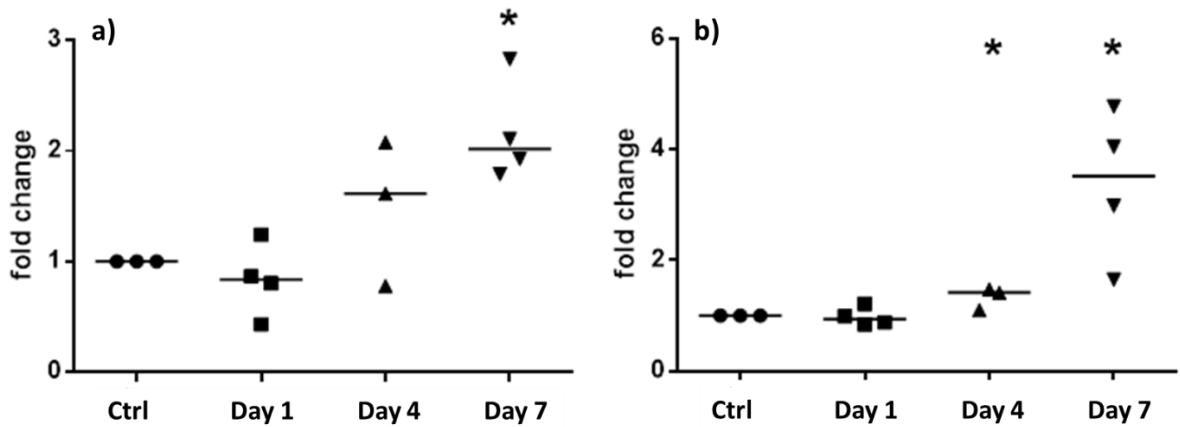

*Figure 9: Fold change of apoptotic cells in a) peripheral blood and b) spleen, following irradiation using the protocol shown in Figure 7. Symbols correspond to individual mice, the center line and error bars are average and Standard deviation. Asterisk denotes p<0.05 compared to control.*

**One month study**

Figure 10 shows the measured dose, dose rate, temperature and humidity during one of the 30 day studies. As the VADER mouse hotel is limited to a capacity of 15 mice, experiments were run separately for the early and late time points. Fig 10 shows a run where time points of 5, 20 and 30 days were assessed.

The measured dose rate, reconstructed from source position, was confirmed by the per-mouse TLD dosimetry and matched the required dose points. Temperature and humidity were within the accepted parameters. The decline in humidity is likely due to removal of 1/3 of the mice on days 5 and 20 (air flow into the mouse hotel was not humidified).

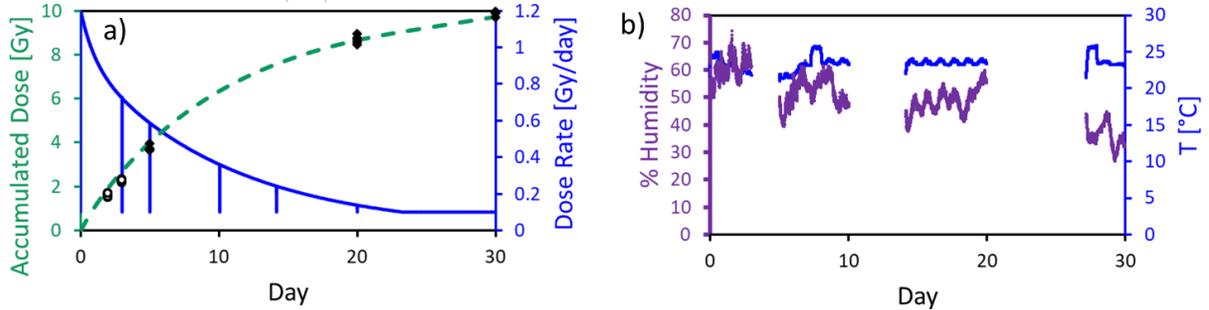

*Figure 10: a) Dose rate (solid line, right axis) and cumulative dose (dashed line, left axis) calculated from source position over a 30-day study. The vertical lines in the dose rate curve correspond to brief VADER openings for cage cleaning and removing mice for analysis. Symbols at days 2, 3, 5, 20 and 30 are individual TLD readings for 5 mice at each time point (open symbols from a separate 5-day study using the same dose profile, conducted immediately prior to the 30-day study). b) Measured humidity (thick purple line, left axis) and temperature (thin blue line, right axis) in one of the cages used for the VADER. The gaps represent times when a different cage was used.*

*Table 2: Dose rates and doses for the 30-day study. The TLD average and standard deviation dose based on 5 mice each. P values are based on an unpaired t test, comparing to controls.*

| Day | Dose rate [Gy/day] | Nominal Dose [Gy] | Average TLD dose [Gy] | Leukocyte counts | p value |
|---|---|---|---|---|---|
| Control | 0 | 0 | -0.05±0.08 | 2300±800 | NA |
| 2 | 0.82 | 1.9 | 1.6±0.1 | 1100±300 | 0.005 |
| 3 | 0.72 | 2.7 | 2.3±0.1 | 1000±800 | 0.007 |
| 5 | 0.59 | 4.0 | 3.7±0.1 | 800±400 | 0.0004 |
| 20 | 0.14 | 8.7 | 8.7±0.2 | 1100±400 | 0.003 |
| 30 | 0.10 | 9.8 | 9.95±0.2 | 1300±600 | 0.002 |

Figure 11 and Table 2 show the leukocyte counts for this study, which declined similarly to those in Figure 8a with the beginning of recovery evident at Day 20-30. The main goal of this experiment was to mimic the injected $^{137}$Cs experiments reported previously [16], indeed, both the cumulative doses and the pattern of up and down regulated genes seen in that study was reproduced in this experiment (manuscript in preparation).

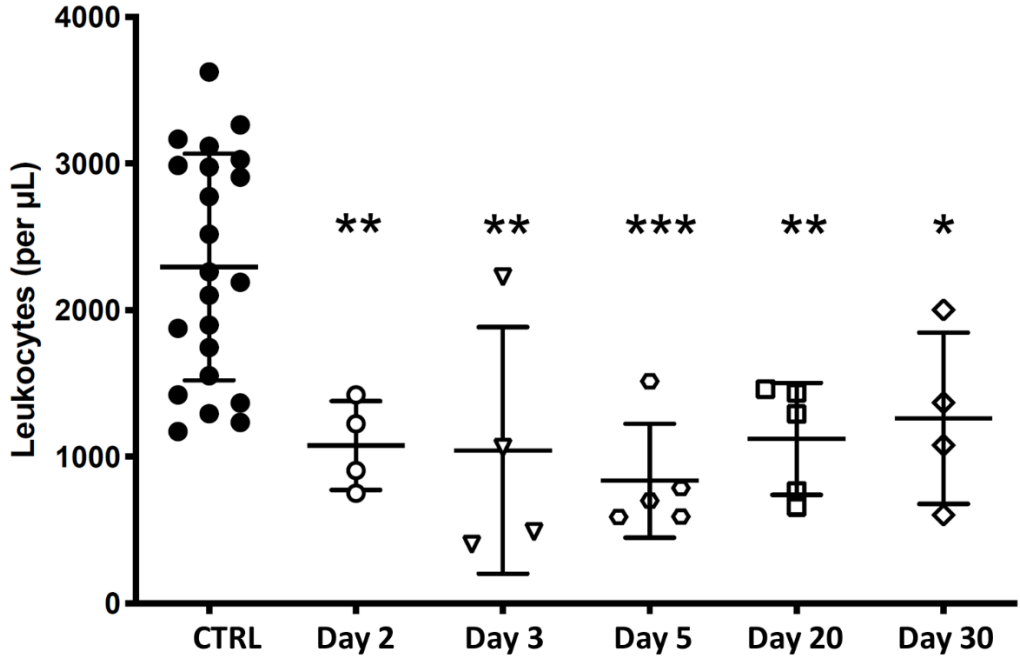

*Figure 11: Leukocyte (CD45+) counts at various times during a 30-day study. Symbols correspond to individual mice, the center line and error bars are average and standard deviation. Single, double and triple asterisks denote p<0.05, p<0.001 and p<0.001, compared to control, respectively.*

## Discussion

We present here a novel low-dose-rate irradiator based on re-purposed $^{137}$Cs brachytherapy seeds. The irradiator was designed to mimic internal or external exposures from high level environmental $^{137}$Cs contamination, resulting in chronic exposures of Gy-level doses over days to weeks of exposure. Thus the VADER was designed to provide dose rates of 0.1-1 Gy/day. Obtaining a wider range of doses can be achieved by building a taller structure, allowing the sources to be retracted further from the mice, but due to the $1/r^2$ decrease of dose rate the system would have to be too tall to fit in a standard room to gain significantly in this respect. Another option, for achieving lower dose rates, if required, is to remove some of the seeds or replace them with weaker ones, essentially scaling down all doses.

**Alternate Technologies**

The Howell group have demonstrated the validity of modeling internal $^{137}$Cs exposures with external exposures using a cabinet type $^{137}$Cs irradiator coupled to a computer controlled mercury attenuator [28, 29]. In their system a mercury reservoir is placed between an 18 Ci $^{137}$Cs source and one or more cages holding four mice each. The amount of mercury in the reservoir can be modified, attenuating the γ-ray flux to achieve dose rates between approximately $10^{-4}$ and 0.25 Gy/h [28]. This setup has been routinely used for calibration studies for internal emitter biomarker studies [30-32].

An alternate approach to attenuation, which may alter the energy spectrum of the source, is to place the source very far from the animals to be irradiated. An example of this approach is the NMBU FIGARO low dose rate facility in Norway [33]. FIGARO utilizes a $^{60}$Co source, providing a dose rate between 1 Gy/h, just outside the collimator and 0.4 mGy/h, 20 meters away. The obvious drawback of such a system is that it requires a very large facility, which is not suited to an urban university/medical center. FIGARO also does not currently provide variable dose rates, though they could be easily achieved by periodically moving the animal cages. As dose rate varies with the distance from the source, care needs to be taken that the animals cannot move too much in this direction. This is not a huge problem when irradiating at a source distance of 20 m but becomes problematic for higher dose rates that require source distances of 1-2 m.

**VADER performance**

*General*

At the time of writing this paper the VADER has been in operation for over 9 months, including two 30-day studies and seven 7-day studies, for the study of cytogenetic, transcriptomic, and metabolomic endpoints. In addition to the design and dosimetry validation of the VADER we presented here blood count and apoptosis data from one 7 day study and one 30-day study. Additional manuscripts focusing on these different endpoints will be presented elsewhere. It should be noted that in our prior injection-based studies we were not able to access hematological data (especially at earlier time points) as the blood was too radioactive to analyze. Use of the VADER obviates this problem.

In the first pilot study (conducted in August 2018) the temperature in the VADER ranged from 26 °C to 30 °C. Temperature in the control cage was slightly lower (23 °C to 29 °C), although both were somewhat warmer than the recommended temperatures for housing mice. Humidity in both hotels ranged from 40% to 60%. Following this week-long experiment, a second air conditioning unit was installed in the room, lowering summer temperatures to more acceptable levels.

During this pilot study we also discovered several errors in data logging, which were corrected. One power outage had been experienced, but the VADER control software had recovered successfully and resumed operation when the power was restored.

*Dosimetry*

An important aspect of performing these long-term studies, where mice are free to roam in a (possibly) spatially varying radiation field is individualized dosimetry. We have developed an *in vivo* dosimetry technique [26], based on glass encapsulated TLD rods. The glass encapsulation allows the TLDs to be injected into mice without being damaged by the surrounding biofluids, while also allowing readout and annealing in the encapsulated state.

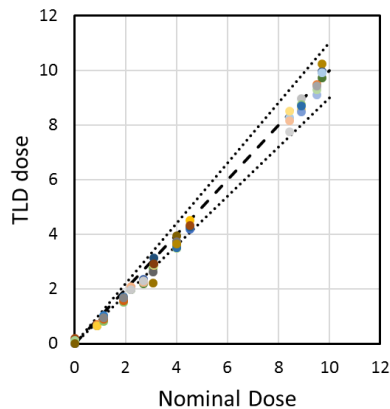

*Figure 12: TLD reconstructed doses (in Gy) for mice irradiated in the VADER. Each point corresponds to an individual mouse dose. The dashed line is the identity and the dotted lines are ±10%.*

Figure 12 demonstrates *in vivo* measured doses, reconstructed using the TLDs, as a function of the nominal (requested) dose. The data show good agreement at the higher doses but the TLD doses were slightly lower than nominal at the lower doses. This may be due to the fact that most low doses were delivered in a short time using the highest dose rate setting, where there is considerable spatial variation of dose at dose rates above 1 Gy/day, and it is possible that mice spent more time in the lower dose areas.

# Conclusions

We describe a low dose irradiation facility based on repurposed $^{137}$Cs brachytherapy seeds. The VADER provides dose rates between 0.1 and 1.2 Gy/day to up to 15 mice at a time. Environmental conditions suitable for housing mice are maintained in the VADER for weeks at a time allowing protracted experiments with short weekly breaks for cage cleaning and food/water replenishment.

The system is currently in routine use at Columbia University and can reproduce the results of $^{137}$Cs injection experiments without the complications and costs involved in the handling of radioactive reagents, animals, and biofluids.

# Funding

This work was supported by the National Institute of Allergy and Infectious Diseases, National Institutes of Health, grant number U19 AI067773 to the Center for High-Throughput Minimally-Invasive Radiation Biodosimetry. The content is solely the responsibility of the authors and does not necessarily represent the official views of National Institute of Allergy and Infectious Diseases or of the National Institutes of Health.